\documentclass[authoryear,12pt,a4paper]{elsarticle}
\usepackage[latin1]{inputenc}
\usepackage{color, colortbl}
\definecolor{OliveGreen}{cmyk}{0.64,0,0.95,0.40}
\usepackage{tikz}
\usetikzlibrary{shapes}
\usetikzlibrary{patterns}
\usepackage{algorithm,algorithmic}
\algsetup{indent=2em}

\usepackage{subfigure}
\usepackage{amsmath}
\usepackage{amssymb}
\usepackage{amsfonts}
\usepackage{amsbsy}
\usepackage{bbm}

\usepackage[OT1]{fontenc}
\usepackage[charter]{mathdesign}
\usepackage[OMLmathbf,sfdefault=cmbr]{isomath}

\newcommand\removed[1]{\marginpar{\raggedright\footnotesize\sf\textcolor{red}{\textbf{Removed}}}\textcolor{red!50}{#1}}
\definecolor{OliveGreen}{cmyk}{0.64,0,0.95,0.40}

\renewcommand\removed[1]{\strut}

\usepackage[pdfborder=0 0 0,colorlinks=true]{hyperref} 
\usepackage{graphicx}
\DeclareGraphicsExtensions{.pdf}

\renewcommand{\eqref}[1]{Eq.~(\ref{#1})}

\newcommand{\cd}{{\mathcal D}}

\newcommand{\cl}{{\mathcal L}}

\newcommand{\cn}{{\mathcal N}}

\newcommand{\cp}{{\mathcal P}}

\newcommand{\cs}{{\mathcal S}}

\newcommand{\cu}{{\mathcal U}}

\newcommand{\cx}{{\mathcal X}}

\newcommand{\cz}{{\mathcal Z}}

\newcommand{\Nn}{{\mathbb N}}

\newcommand{\Rr}{{\mathbb R}}

\newcommand{\ldeu}[2]{\cl_2\left(#1,\,#2\right)}

\newcommand{\di}[1]{{\rm d}#1} 				
\newcommand{\ve}[1]{\boldsymbol{#1}}			
\newcommand{\ma}[1]{\boldsymbol{\rm #1}}		

\newcommand{\tr}{^{\textsf T}}				

\newcommand{\enu}{ , \, \dots \,,}

\newcommand{\acc}[1]{\left\{#1\right\}}			

\newcommand{\Var}[1]{{\rm Var}\left[ #1 \right]}

\newcommand{\Esp}[1]{{\mathbb E}\left[ #1 \right]}
\newcommand{\Espe}[2]{{\mathbb E}_{#1}\left[#2\right]}

\newcommand{\Prob}[1]{{\mathbb P}\left( #1 \right)}	

\newcommand{\ie}{{\em i.e.} }
\newcommand{\eg}{{\em e.g.} }

\usepackage{multirow}                       
\makeatletter
\def\hlinewd#1{
\noalign{\ifnum0=`}\fi\hrule \@height #1 %
\futurelet\reserved@a\@xhline}
\makeatother
\newcommand{\hlineT}{\hlinewd{1.1pt}}
\newcommand{\hlineB}{\hlinewd{0.85pt}}

\journal{Probabilistic Engineering Mechanics}

\begin{document}


\begin{frontmatter}

    \title{Metamodel-based importance sampling for structural reliability analysis}
    \author[LaMI,Phimeca]{V. Dubourg\corref{cor1}}
    \ead{dubourg@phimeca.com}
    \author[Phimeca]{F. Deheeger}
    \ead{deheeger@phimeca.com}
    \author[LaMI,Phimeca]{B. Sudret}
    \ead{sudret@phimeca.com}

    \cortext[cor1]{Corresponding author}
    \address[LaMI]{Clermont Universit\'e, IFMA, EA 3867, Laboratoire de M\'ecanique et Ing\'enieries, BP 10448, F-63000 Clermont-Ferrand}
    \address[Phimeca]{Phimeca Engineering, Centre d'Affaires du Z\'enith, 34 rue de Sarli\`eve, F-63800 Cournon d'Auvergne}

    \begin{abstract}
        Structural reliability methods aim at computing the probability of failure of systems with respect to some prescribed performance functions. In modern engineering such functions usually resort to running an expensive-to-evaluate computational model (\eg a finite element model). In this respect simulation methods, which may require $10^{3-6}$ runs cannot be used directly. {\em Surrogate models} such as quadratic response surfaces, polynomial chaos expansions or kriging (which are built from a limited number of runs of the original model) are then introduced as a substitute of the original model to cope with the computational cost. In practice it is almost impossible to quantify the error made by this substitution though. In this paper we propose to use a kriging surrogate of the performance function as a means to build a quasi-optimal importance sampling density. The probability of failure is eventually obtained as the product of an {\em augmented} probability computed by substituting the meta-model for the original performance function and a {\em correction term} which ensures that there is no bias in the estimation even if the meta-model is not fully accurate. The approach is applied to analytical and finite element reliability problems and proves efficient up to 100 random variables.
    \end{abstract}

    \begin{keyword}
        reliability analysis \sep importance sampling \sep metamodeling error \sep kriging \sep random fields \sep active learning \sep rare events
    \end{keyword}

\end{frontmatter}


\section{Introduction}

Reliability analysis consists in the assessment of the level of safety of a system. Given a probabilistic model (an $n$-dimensional random vector $\ve{X}$ with \emph{probability density function} (PDF) $f_{\ve{X}}$) and a performance model (a function $g$), it makes use of mathematical techniques in order to estimate the safety level of the system in the form of a failure probability. A basic reference approach is the Monte Carlo simulation technique that resorts to numerical simulation of the performance function through the probabilistic model.\par

Failure is usually defined as the event $F=\acc{g\left(\ve{X}\right) \leq 0}$, so that the failure probability is defined as follows:%
\begin{equation} \label{eq:pf_def}
    \begin{split}
        p_f & \equiv \Prob{F} = \Prob{\acc{g\left(\ve{X}\right) \leq 0}} \\
            & = \int_{\cd_f = \acc{\ve{x}\in\Rr^n\,:\,g\left(\ve{x}\right) \leq 0}} f_{\ve{X}}\left(\ve{x}\right)\,\di{\ve{x}}
    \end{split}
\end{equation}%
Introducing the failure indicator function $\mathbbm{1}_{g \leq 0}$ being equal to one if $g\left(\ve{x}\right) \leq 0$ and zero otherwise, the failure probability turns out to be the mathematical expectation of this indicator function with respect to the joint probability density function $f_{\ve{X}}$ of the random vector $\ve{X}$. This convenient definition allows one to derive the Monte Carlo estimator which reads:%
\begin{equation}
  \begin{split}
    \widehat{p}_{f\,{\rm MC}} &\equiv \widehat{\mathbb{E}}_{\ve{X}}\left[\mathbbm{1}_{g \leq 0}\left(\ve{X}\right)\right] \\
			      &= \frac{1}{N}\,\sum_{k=1}^N \mathbbm{1}_{g \leq 0}\left(\ve{x}^{\left(k\right)}\right)
  \end{split}
\end{equation}%
where $\acc{\ve{x}^{\left(1\right)}\enu \ve{x}^{\left(N\right)}}$, $N\in\Nn^*$ is a set of samples from the random vector $\ve{X}$. According to the central limit theorem, this estimator is asymptotically unbiased and normally distributed with variance:%
\begin{equation}
 \sigma_{\widehat{p}_{f\,{\rm MC}}}^2 \equiv \Var{\widehat{p}_{f\,{\rm MC}}} = \frac{p_f\,\left(1-p_f\right)}{N}
\end{equation}%
In practice, this variance is compared to the unbiased estimate of the failure probability in order to decide whether the simulation is accurate enough or not. To do so, one usually resorts to the coefficient of variation defined as $\delta_{\widehat{p}_{f\,{\rm MC}}} = \sigma_{\widehat{p}_{f\,{\rm MC}}}/\widehat{p}_{f\,{\rm MC}}$. $N$ being fixed, this coefficient of variation dramatically increases when the failure event becomes rare ($p_f \rightarrow 0$). This makes the Monte Carlo estimation technique intractable for real world engineering problems for which the performance function involves the output of an expensive-to-evaluate black-box function -- \eg a finite element code. Note that this remark is also true for too frequent events ($p_f \rightarrow 1$) as one should compute the coefficient of variation of $1-\widehat{p}_{f\,{\rm MC}}$ that exhibits the same property.\par

In order to reduce the number of simulation runs, different alternatives to the brute-force Monte Carlo method have been proposed and might be classified as follows.\par

One first approach consists in replacing the original experiment by a \emph{surrogate} which is much faster to evaluate. Various surrogates have been used amongst which are: quadratic response surfaces \citep{Bucher1990,Kim,Das} and the more recent metamodels such as \emph{support vector machines} \citep{Hurtado2004b,Deheeger2007}, \emph{neural networks} \citep{Papadrakakis2002} and \emph{kriging} \citep{Kaymaz2005,Bichon2008}. Nevertheless, it is often difficult or even impossible to quantify the error made by such a substitution.\par

The well-known first- and second-order reliability methods, based on Taylor expansions somewhat differ from these surrogate-based approaches because of their mathematical background \citep{Breitung84}. But in practice they feature the same limitation: the assumptions (mostly the unicity of the so-called \emph{most probable failure point}) they are built on may not hold and it is difficult to validate them.

\citet{Bucher2009,Naess2009,Nishijima2010} investigate the use of the \emph{extreme value theory} in order to infer the tail of the \emph{cumulative distribution function} (CDF) of the random variable $g(\ve{X})$ from a set of samples. This set is obtained by crude Monte Carlo simulation of the input random vector $\ve{X}$ propagated through the performance function $g$. Despite this method has a lot of attractive features (it is independent of both the dimension $n$ and the shape of $g$), it may require a large number of samples to make the fitted \emph{extreme value distribution} accurate far in the tail -- \ie for low failure probabilities.\par

From another point of view, \emph{variance reduction techniques} have been proposed in order to make Monte Carlo simulation more efficient. Importance sampling \citep{Rubinstein2008} aims at concentrating the Monte Carlo samples in the vicinity of the limit-state surface, \eg around the most probable failure point (also known as \emph{design point}) obtained by a preliminary FORM analysis \citep{Melchers1989}. Subset simulation \citep{Au2001,Ching2005b,Ching2005,Santoso2010} computes the failure probability as a product of conditional probabilities, each of them being estimated by Markov Chain Monte Carlo simulation. All these approaches reveal robust, although they are often too much computationally demanding to be implemented in industrial cases.\par

As a summary the current practice for evaluating probabilities of failure in case of computationally demanding performance functions relies on the substitution of the limit-state function by a metamodel for which no general error estimation is usually available.\par

In this paper, a new hybrid approach combining importance sampling and an adaptive metamodeling technique is proposed. First, a kriging surrogate of the limit-state function is built and adaptively refined. Then, the probabilistic prediction provided by the kriging surrogate is used to build up a quasi-optimal importance sampling density. As a result the probability of failure is computed as the product of two terms, one which is estimated by sampling the surrogate limit-state function, and a correction factor computed from the original limit-state function.\par

The paper is organized as follows. Section \ref{sec:metamodeling} recalls the basic of Gaussian process (kriging) metamodeling and introduces the probabilistic classification function. Section \ref{sec:metaIS} presents the construction of a quasi-optimal importance sampling density derived from this function and the associated estimator of the failure probability. Section \ref{sec:DOE} introduces an adaptive refinement technique of this importance sampling density function so as to make the algorithm as parsimonious as possible. Section \ref{sec:Implementation} and \ref{sec:Appli} eventually describe practical implementation details and application examples.\par


\section{Probabilistic classification using \emph{margin metamodels}} \label{sec:metamodeling}

A metamodel means to a model what the model itself means to the real-world. Loosely speaking, it is \emph{the model of the model}. As opposed to the model, its construction does not rely on any physical assumption about the phenomenon of interest but rather on statistical considerations about the coherence of some scattered observations that result from a set of experiments. This set is usually referred to as a \emph{design of experiments} (DOE): $\cx = \{\ve{x}_1\enu\ve{x}_m\}$. It should be carefully selected in order to retrieve the largest amount of statistical information about the underlying functional relationship over the input space $\cd_{\ve{x}}$. In the sequel one attempts to build a metamodel for the failure indicator function $\mathbbm{1}_{g \leq 0}$. In the \emph{statistical learning theory} \citep{Vapnik1995} this is referred to as a \emph{classification} problem.\par

In this paper, we define a \emph{margin metamodel} as a metamodel that is able to provide a \emph{probabilistic prediction} of the response quantity of interest whose spread (\ie variance) depends on the amount of information brought by the DOE. Note that this spread is thus reducible by bringing more observations into the DOE. In other words, this is an \emph{epistemic} (reducible) source of uncertainty that shall not be confused with the possible uncertainty in $\ve{X}$ that will be modelled by a PDF $f_{\ve{X}}$ later on for structural reliability purposes. To the authors' knowledge, there exist only two families of such margin metamodels: the probabilistic support vector machines $\mathbb{P}$-SVM by \citet{Platt1999} and Gaussian process- (or kriging-) based classification. The present paper makes use of the kriging metamodel, but the overall concept could easily be extended to $\mathbb{P}$-SVM. The theoretical aspects of the kriging prediction are briefly introduced in the following subsection before it is applied to the classification problem of interest.\par

\subsection{Gaussian-process based prediction}

The Gaussian process (also known as \emph{kriging}) metamodeling theory is detailed in \citet{Sacks89,Welch1992,Santner2003}. In essence, it assumes that the performance function $g$ is a sample path of an underlying Gaussian stochastic process $G$ that may be cast as follows:%
\begin{equation}
    G\left(\ve{x}\right) = \ve{f}\left(\ve{x}\right)\tr\,\ve{\beta} + Z\left(\ve{x}\right)
\end{equation}%
where:%
\begin{itemize}
    \item $\ve{f}\left(\ve{x}\right)\tr\,\ve{\beta}$ denotes the mean of the GP which corresponds to a classical linear regression model on a given functional basis $\acc{f_i,\;i = 1\enu p} \in \ldeu{\cd_{\ve{x}}}{\Rr}$~;
    \item $Z\left(\ve{x}\right)$ denotes a zero-mean stationary Gaussian process with a constant variance $\sigma_G^2$. It is fully defined by its autocovariance function which reads:%
    \begin{equation}
        C_{GG}\left(\ve{x},\,\ve{x}'\right) = \sigma_G^2\,R\left(\ve{x}-\ve{x}',\ve{\ell}\right)
    \end{equation}%
    where $\ve{\ell}$ is a vector of parameters defining $R$.
\end{itemize}%
The most widely used class of autocorrelation functions is the \emph{anisotropic squared exponential} model:%
\begin{equation} \label{eq:exp2}
    R\left(\ve{x}-\ve{x}',\,\ve{\ell}\right) = \exp\left(- \sum\limits_{k=1}^n \left(\frac{x_k-x_k'}{\ell_k}\right)^2\right)
\end{equation}%

The \emph{best linear unbiased estimation} (BLUE) of $G$ at point $\ve{x}$ is shown \citep{Santner2003} to be the following Gaussian random variate:
\begin{equation} \label{eq:BLUE}
    \begin{split}
        \widehat{G}\left(\ve{x}\right) & = G(\ve{x}) \left| \{G(\ve{x}_1) = g(\ve{x}_1)\enu G(\ve{x}_m) = g(\ve{x}_m)\} \right. \\
                                       & \sim \cn\left(\mu_{\widehat{G}}\left(\ve{x}\right),\,\sigma_{\widehat{G}}\left(\ve{x}\right)\right)
    \end{split}
\end{equation}
with moments:%
\begin{align}
    \mu_{\widehat{G}}\left(\ve{x}\right) & = \ve{f}\left(\ve{x}\right)\tr\,\ve{\widehat{\beta}} + \ve{r}\left(\ve{x}\right)\tr\ma{R}^{-1}\left(\ve{y} - \ma{F}\,\ve{\widehat{\beta}}\right) \\
    \sigma_{\widehat{G}}^2\left(\ve{x}\right) & = \sigma_{G}^2\,\left(
    1
    - \ve{r}(\ve{x})\tr\,\ma{R}^{-1}\,\ve{r}(\ve{x})
    + \ve{u}(\ve{x})\tr\,(\ma{F}\tr\,\ma{R}^{-1}\,\ma{F})^{-1}\,\ve{u}(\ve{x})
    \right)
\end{align}%
where $\ve{y} = \langle g(\ve{x}_1) \enu g(\ve{x}_m) \rangle\tr$ is the vector of observations, $\ma{R}$ is their correlation matrix, $\ve{r}(\ve{x})$ is the vector of cross-correlations between the observations and the prediction, and $\ma{F}$ is the so-called regression matrix. The terms of $\ma{R}$, $\ve{r}(\ve{x})$ and $\ma{F}$ are respectively defined as follows:%
\begin{align}
    r_i(\ve{x}) & = R\left(\ve{x}-\ve{x}_i,\,\ve{\ell}\right),\;i=1\enu m \\
    R_{ij} & = R\left(\ve{x}_i-\ve{x}_j,\,\ve{\ell}\right),\;i,\,j =1\enu m\\
    F_{ij} & = f_j\left(\ve{x}_i\right),\;i=1\enu m,\;j=1\enu p
\end{align}
Finally, the generalized least-squares solution $\widehat{\ve{\beta}}$ and the vector $\ve{u}(\ve{x})$ respectively read:
\begin{align}
    \widehat{\ve{\beta}} & = (\ma{F}\tr\,\ma{R}^{-1}\,\ma{F})^{-1}\,\ma{F}\tr\,\ma{R}^{-1}\,\ve{y} \\
    \ve{u}(\ve{x}) & = \ma{F}\tr\,\ma{R}^{-1}\,\ve{r}(\ve{x}) - \ve{f}(\ve{x})
\end{align}%
At this stage one can easily prove that $\mu_{\widehat{G}}\left(\ve{x}_i\right) = g\left(\ve{x}_i\right)$ and $\sigma_{\widehat{G}}\left(\ve{x}_i\right) = 0$ for $i = 1\enu m$, thus meaning the kriging surrogate \emph{interpolates} the observations without any residual epistemic uncertainty.

Given a choice for the regression and correlation models, the optimal set of parameters $\ve{\beta}^*$, $\ve{\ell}^*$ and $\sigma_{G}^{2\,*}$ can then be inferred using the \emph{maximum likelihood principle} applied to the single sparse observation of the GP sample path grouped into the vector $\ve{y}$. This inference problem turns into an optimization problem that can be solved analytically for both $\ve{\beta}^*$ and $\sigma_{G}^{2\,*}$ assuming $\ve{\ell}^*$ is known. Thus the problem is solved in two steps: the maximum likelihood estimation of $\ve{\ell}^*$ is first solved by a global optimization algorithm which in turns allows one to evaluate the optimal $\ve{\beta}^*$ and $\sigma_{G}^{2\,*}$. Implementation details can be found in \citet{Welch1992,Lophaven2002}.\par

\subsection{Probabilistic classification function} \label{sec:ProbabilisticClassificationFunction}

As seen from \eqref{eq:BLUE}, the kriging surrogate provides both an approximation of the limit-state function $g(\ve{x})$ which is denoted by $\mu_{\widehat{G}}(\ve{x})$ and an epistemic prediction uncertainty which is characterized by the kriging variance $\sigma^2_{\widehat{G}}(\ve{x})$.\par

Authors usually make use only of the surrogate $\widetilde{g}(\ve{x}) \equiv \mu_{\widehat{G}}(\ve{x})$ for reliability analysis -- see \citet{Kaymaz2005,Bichon2008}. In this paper it is proposed to use the complete probabilistic prediction -- see \citet{Picheny2009} for a similar idea.\par

Let us introduce for this purpose the following \emph{probabilistic classification function}:%
\begin{equation}
    \pi(\ve{x}) \equiv \cp\left[\widehat{G}(\ve{x}) \leq 0\right]
\end{equation}%
In this expression, the probability measure $\cp\left[\bullet\right]$ refers to the Gaussian nature of the kriging predictor $\widehat{G}(\ve{x}) \sim \cn(\mu_{\widehat{G}}(\ve{x}),\,\sigma_{\widehat{G}}(\ve{x}))$ and shall not be confused with the probability measure $\Prob{\bullet}$ associated with the random vector $\ve{X}$ -- see \eqref{eq:pf_def}.\par

Thanks to the Gaussian nature of the kriging predictor, the probabilistic classification function rewrites:%
\begin{equation} \label{eq:ProbabilisticClassificationFunction}
    \pi(\ve{x}) = \Phi\left(\frac{0-\mu_{\widehat{G}}(\ve{x})}{\sigma_{\widehat{G}}(\ve{x})}\right) \; \text{if} \; \ve{x} \notin \cx
\end{equation}%
and still holds for the points in the experimental design for which the prediction variance equals zero by switching to the limit:
\begin{equation}
    \pi(\ve{x}) = \left\{\begin{array}{rcl}
                    1 & \text{if} & \ve{x} \in \cx, \; g(\ve{x}) \leq 0 \\
                    0 & \text{if} & \ve{x} \in \cx, \; g(\ve{x}) > 0
                  \end{array}\right.
\end{equation}
It shall be again emphasized that $\pi(\ve{x})$ is \emph{not} the sought failure probability. It may be interpreted as the probability that the predictor $\widehat{G}(\ve{x})$ (for some prescribed deterministic $\ve{x}$) is negative with respect to the epistemic uncertainty.\par

\begin{figure}
    \centering
    \includegraphics[width=.75\linewidth, clip=true, trim=30 -15 30 30]{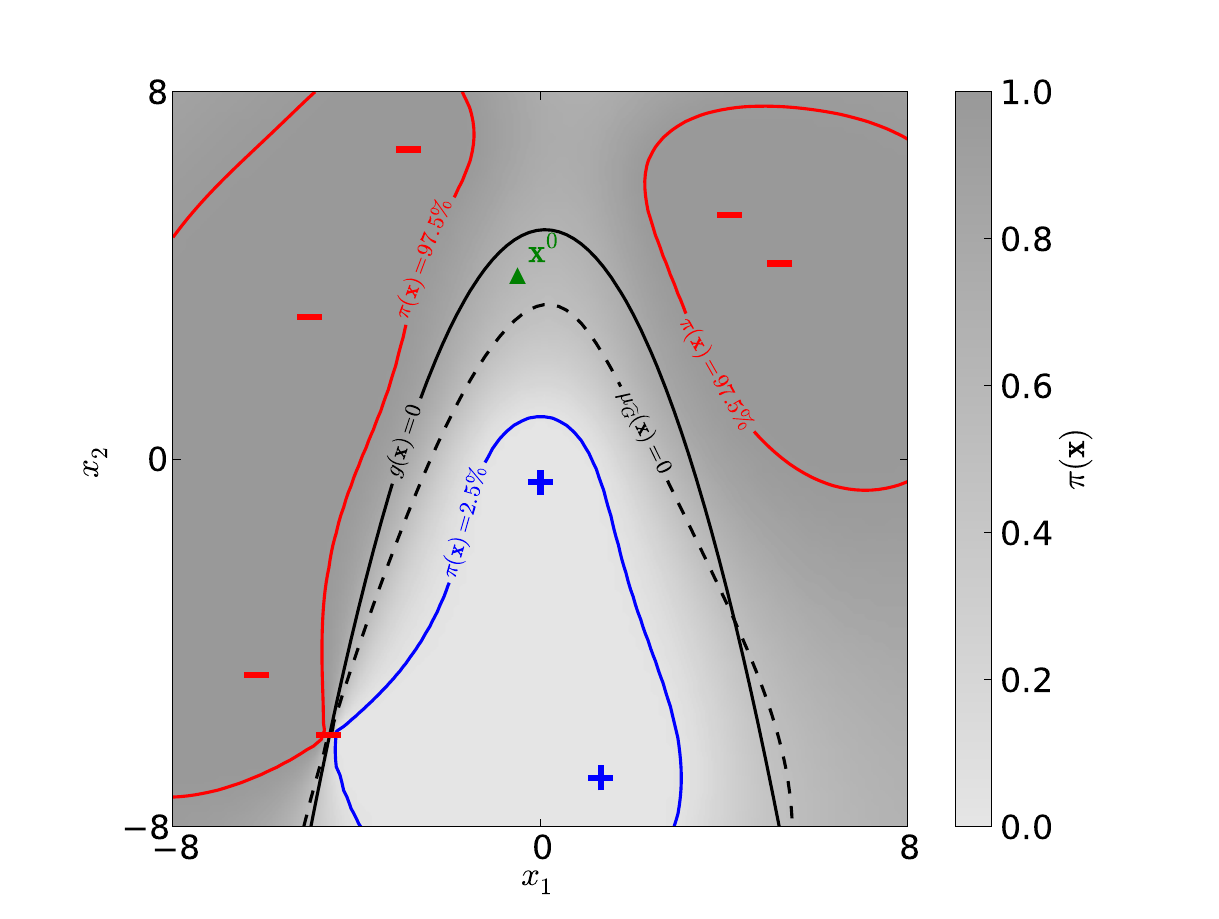}
    \caption{Comparison of the two classification strategies on a two-dimensional example from \citet{DerKiureghian1998} -- Limit-state functions and kriging probability values.}
    \label{fig:ProbabilisticClassificationFunction_2D}
\end{figure}

Figure \ref{fig:ProbabilisticClassificationFunction_2D} illustrates the concepts introduced in this section on a basic structural reliability example from \citet{DerKiureghian1998}. This example involves two independent standard Gaussian random variates $X_1$ and $X_2$, and the performance function reads:
\begin{equation}
 g(x_1,\,x_2) = b - x_2 - \kappa\,(x_1 - e)^2
\end{equation}
where $b = 5$, $\kappa = 0.5$ and $e = 0.1$. In Figure \ref{fig:ProbabilisticClassificationFunction_2D}, the original limit-state function $g(\ve{x})=0$ is represented by the black solid line. The red ``$+$'' and the blue ``$-$'' represent the initial DOE from which the kriging metamodel is built, the color being related to the sign of the performance function $g$. Once kriging has been applied, the mean prediction of the limit-state surface, which is defined as $\acc{\ve{x}\in\Rr^n\,:\,\mu_{\widehat{G}}(\ve{x}) = 0}$, is represented by the dashed black line. It can be seen that the metamodel is not fully accurate since the green triangle $\ve{x}^0$ (among others) is misclassified. Indeed, $\ve{x}^0$ is in the safe domain according to the real performance function $g$, and in the failure domain according to the mean kriging predictor $\mu_{\widehat{G}}$.\par

Note that the probabilistic classification function allows a smoother decision: $\ve{x}^0$ belongs to the failure domain with a 60\% probability w.r.t. the epistemic uncertainty in the random prediction $\widehat{G}(\ve{x}^0)\sim\cn(\mu_{\widehat{G}}(\ve{x}^0),\,\sigma_{\widehat{G}}(\ve{x}^0))$. Note also that the red ``$+$'' and the blue ``$-$'' in the DOE are always surely classified due to the interpolating property of the kriging metamodel ($g(\ve{x}_i) = \mu_{\widehat{G}}(\ve{x}_i),\,i = 1 \enu m$).\par

\begin{figure}
    \centering
    \includegraphics[width=.75\linewidth, clip=true, trim=0 0 0 0]{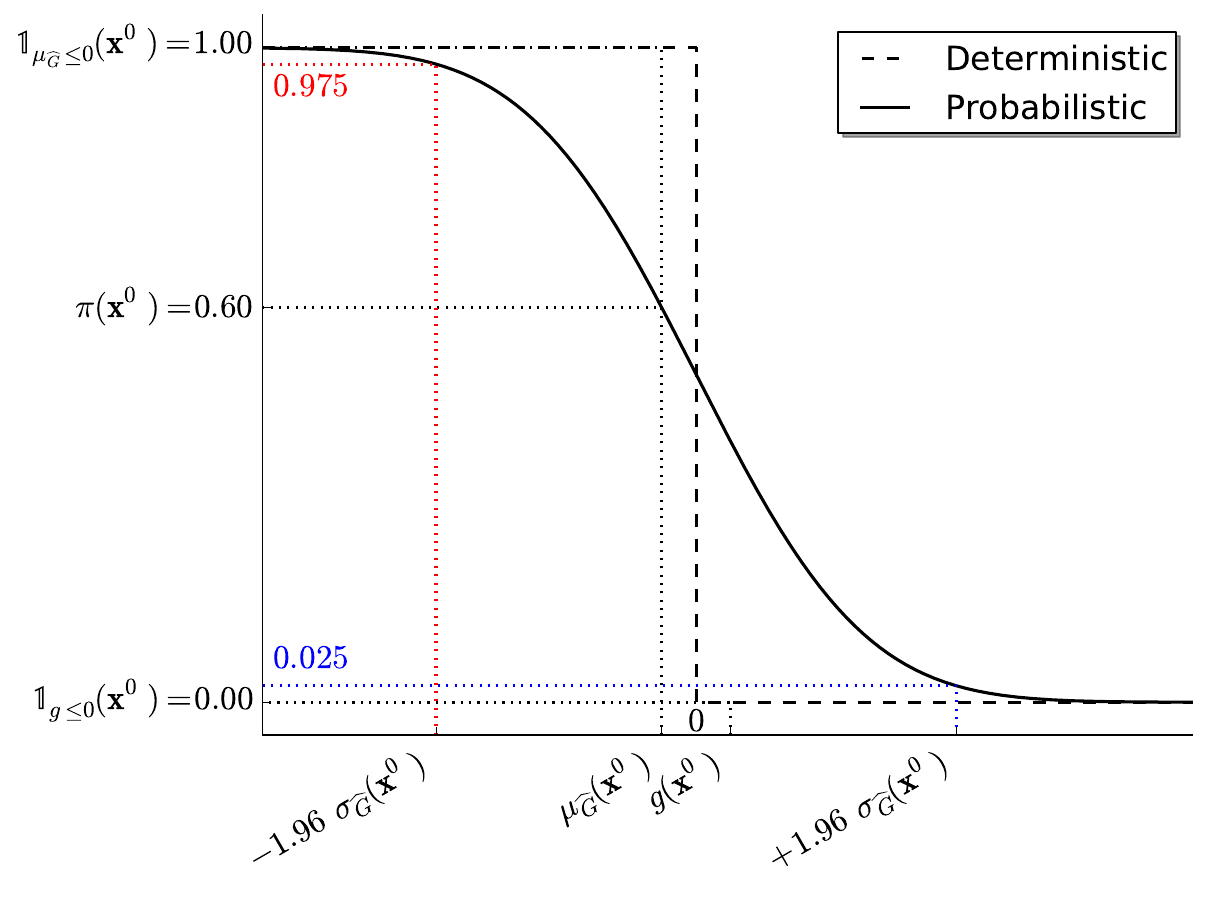}
    \caption{Comparison of the two classification strategies on a two-dimensional example from \citet{DerKiureghian1998} -- Classification functions at point $\ve{x}^0$.}
    \label{fig:ProbabilisticClassificationFunction_1D}
\end{figure}

Figure \ref{fig:ProbabilisticClassificationFunction_1D} is the one-dimensional illustration of the two classification strategies at point $\ve{x}^0$ (green triangle):
\begin{itemize}
    \item according to the deterministic decision function (the Heaviside function centered in zero), the real performance function $g$ is positive whereas the mean kriging prediction $\mu_{\widehat{G}}$ is negative;
    \item according to the probabilistic classification function (the smoother Gaussian cumulative distribution function centered in zero and whose spread is characterized by $\sigma_{\widehat{G}}(\ve{x}^0)$), the epistemic prediction $\widehat{G}$ is negative with a $60\%$ probability ($\pi(\ve{x}^0) = 0.6$ as defined from \eqref{eq:ProbabilisticClassificationFunction}).
\end{itemize}

\section{Metamodel-based importance sampling}\label{sec:metaIS}

\citet{Picheny2009} proposes to use the probabilistic classification function $\pi$ (see \eqref{eq:ProbabilisticClassificationFunction}) as a surrogate for the original indicator function $\mathbbm{1}_{g \leq 0}$, so that the failure probability is rewritten from its definition in \eqref{eq:pf_def} as follows:
\begin{equation}\label{eq:AugmentedFailureProbability}
  \begin{split}
    p_{f\,\varepsilon} &\equiv \int_{\Rr^n} \pi(\ve{x}) f_{\ve{X}}(\ve{x})\,\di{\ve{x}} \\
		       &\equiv \mathbb{E}_{{\ve{X}}}\left[\pi(\ve{X})\right]
  \end{split}
\end{equation}\par

It is argued here that this latter quantity does not equal the failure probability of interest because it sums the aleatory uncertainty in the random vector $\ve{X}$ and the epistemic uncertainty in the prediction $\widehat{G}$. This is the reason why $p_{f\,\varepsilon}$ will be referred to as the \emph{augmented failure probability} in the sequel. As a matter of fact, even if the epistemic uncertainty in the prediction can be reduced (\eg by enriching the DOE as proposed in Section \ref{sec:DOE}), it is impossible to quantify the contribution of each source of uncertainty in $p_{f\,\varepsilon}$.\par

This remark motivates the approach introduced in this section where the probabilistic classification function is used in conjunction with the importance sampling technique in order to build a new estimator of the failure probability.\par

\subsection{Importance sampling}

According to \citet{Rubinstein2008}, importance sampling (IS) is one of the most efficient variance reduction techniques. This technique consists in computing the mathematical expectation of the failure indicator function according to a biased PDF which favors the failure event of interest. This PDF is called the \emph{instrumental density}. Let $h$ denote such a probability density such that $h$ dominates $\mathbbm{1}_{g\leq0}\,f_{\ve{X}}$, meaning that:
\begin{equation}
  \forall \ve{x}\in\cd_{\ve{x}},\; h(\ve{x}) = 0 \Rightarrow \mathbbm{1}_{g\leq0}(\ve{x})\,f_{\ve{X}}(\ve{x})=0
\end{equation}\par

Given this instrumental density, the definition of the failure probability of \eqref{eq:pf_def} may be rewritten as follows:
\begin{equation}\label{eq:pfIS_def}
  \begin{split}
    p_f &= \int_{\Rr^n} \mathbbm{1}_{g\leq0}(\ve{x})\,\frac{f_{\ve{X}}(\ve{x})}{h(\ve{x})}\,h(\ve{x})\,\di{\ve{x}} \\
	&\equiv \Espe{h}{\mathbbm{1}_{g \leq 0}(\ve{X}) \frac{f_{\ve{X}}(\ve{X})}{h(\ve{X})}}
  \end{split}
\end{equation}
In this expression, the expectation is now computed with respect to the instrumental density $h$.\par

The latter definition of the failure probability easily leads to the definition of the \emph{importance sampling estimator}:
\begin{equation}
    \widehat{p}_{f\,{\rm IS}} \equiv \frac{1}{N}\,\sum_{k=1}^N \mathbbm{1}_{g \leq 0}\left(\ve{x}^{\left(k\right)}\right)\,\frac{f_{\ve{X}}(\ve{x}^{\left(k\right)})}{h(\ve{x}^{\left(k\right)})}
\end{equation}
where $\acc{\ve{x}^{\left(1\right)}\enu \ve{x}^{\left(N\right)}}$, $N\in\Nn^*$ is a set of samples drawn from the instrumental density $h$. According to the central limit theorem, this estimation is unbiased and its quality may be measured by means of its variance of estimation which reads:
\begin{equation} \label{eq:varISest}
    \Var{\widehat{p}_{f\,{\rm IS}}} = \frac{1}{N-1}\,\left(\frac{1}{N}\,\sum\limits_{k=1}^N \mathbbm{1}_{g \leq 0}(\ve{x}^{(k)})\,\frac{f(\ve{x}^{(k)})^2}{h(\ve{x}^{(k)})^2} -  \widehat{p}_{f\,{\rm IS}}^2\right)
\end{equation}\par

\citet{Rubinstein2008} show that this variance is zero (optimality of the IS estimator) when the instrumental PDF is chosen as:
\begin{equation} \label{eq:optISdens}
  \begin{split}
    h^*(\ve{x}) &= \frac{\mathbbm{1}_{g \leq 0}(\ve{x})\,f(\ve{x})}{\int \mathbbm{1}_{g \leq 0}(\ve{x})\,f(\ve{x})\,\di{\ve{x}}} \\
		&= \frac{\mathbbm{1}_{g \leq 0}(\ve{x})\,f(\ve{x})}{p_f}
  \end{split}
\end{equation}
However this instrumental PDF is not implementable in practice because it involves the sought failure probability $p_f$ in its denominator. The art of importance sampling consists in building an instrumental density which is quasi-optimal.\par

\subsection{A metamodel-based approximation of the optimal instrumental PDF}

Different strategies have been proposed in order to build quasi-optimal instrumental PDF suited for specific estimation problems. For instance, \citet{Melchers1989} uses a standard normal PDF centered onto the \textit{design point} obtained by FORM in the space of the independent standardized random variables. This approach might be inaccurate if the design point is not unique though. \citet{Cannamela2008} use a kriging prediction of the performance function $g$ in order to build an instrumental PDF suited for the estimation of extreme quantiles of the random variate $g(\ve{X})$. However this approach is not applied to the estimation of failure probabilities.\par

\begin{figure}
  \subfigure[\label{fig:OptimalInstrumentalPDF}The optimal instrumental PDF $h^*$.]{\includegraphics[width=.5\linewidth, clip=true, trim=30 25 30 30]{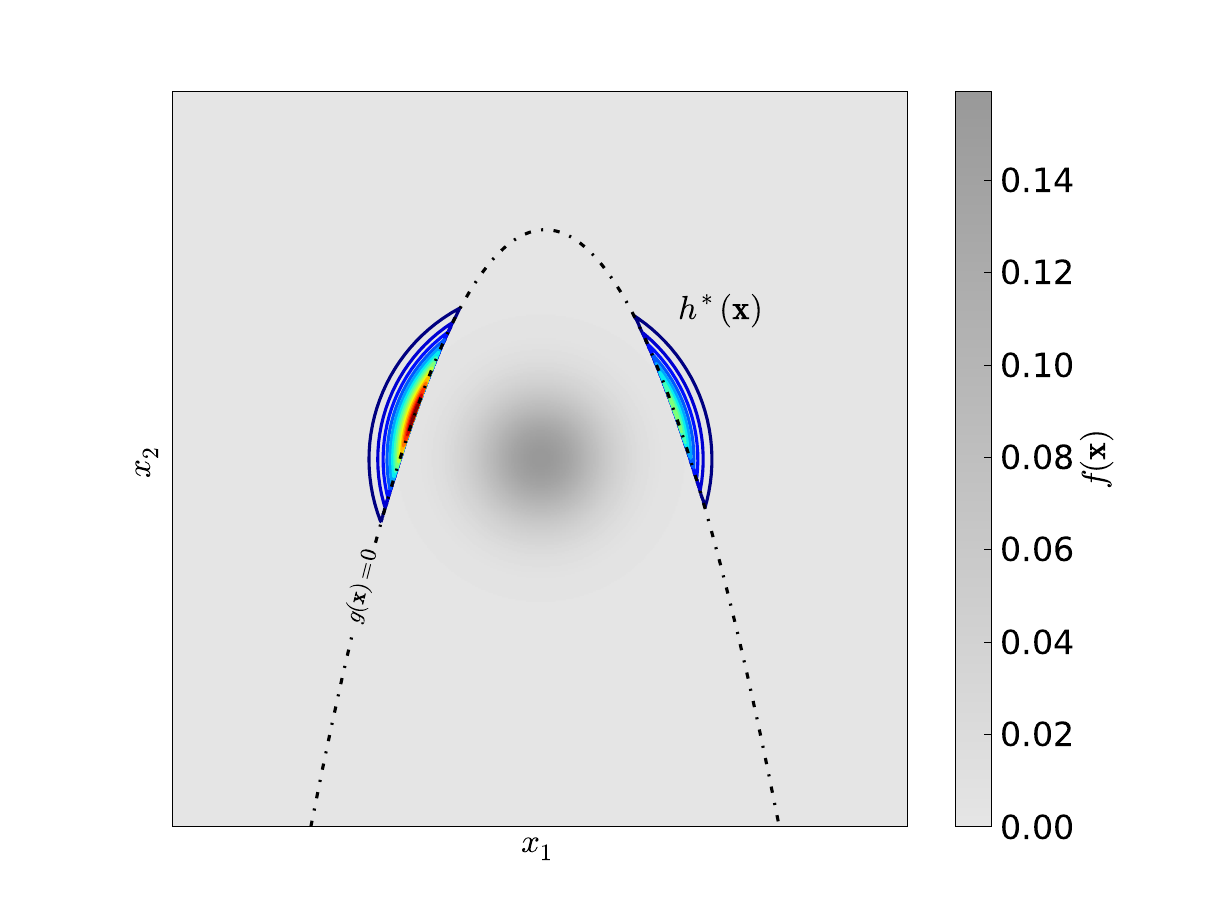}}
  \subfigure[\label{fig:QuasiOptimalInstrumentalPDF}The proposed quasi-optimal instrumental PDF $\widehat{h^*}$.]{\includegraphics[width=.5\linewidth, clip=true, trim=30 25 30 30]{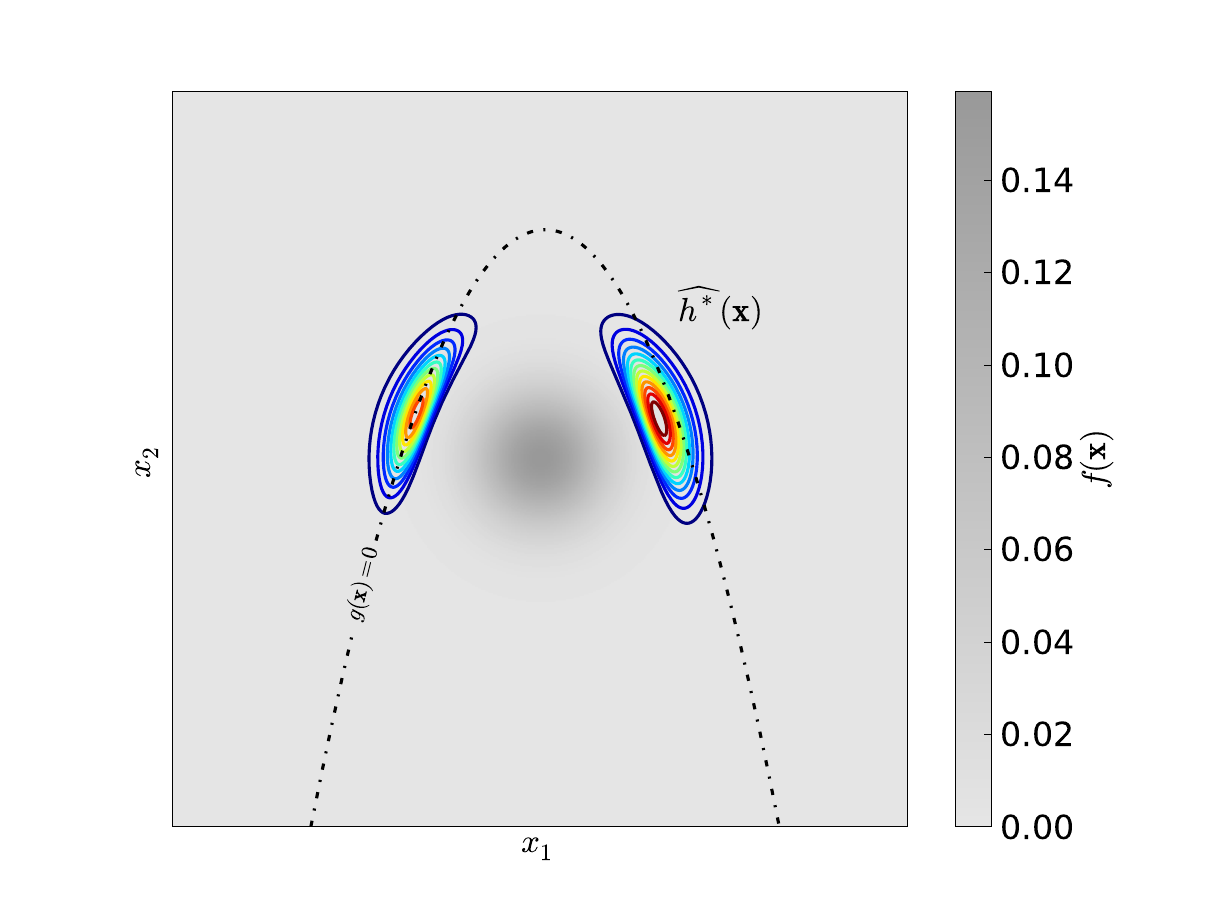}}
  \caption{Comparison of the instrumental PDFs on the two-dimensional example from \citet{DerKiureghian1998}.}
  \label{fig:InstrumentalPDFComparison}
\end{figure}

In this paper it is proposed to use the probabilistic classification function in \eqref{eq:ProbabilisticClassificationFunction} as a surrogate for the real indicator function in the optimal instrumental PDF in \eqref{eq:optISdens}. The proposed quasi-optimal PDF thus reads as follows:
\begin{equation} \label{eq:quasioptISdens}
  \begin{split}
    \widehat{h^{*}}(\ve{x}) &= \frac{\pi(\ve{x})\,f(\ve{x})}{\int \pi(\ve{x})\,f(\ve{x})\,\di{\ve{x}}} \\
			    &\equiv \frac{\pi(\ve{x})\,f(\ve{x})}{p_{f\,\varepsilon}}
  \end{split}
\end{equation}
where $p_{f\,\varepsilon}$ is the augmented failure probability which has been already defined in \eqref{eq:AugmentedFailureProbability}. For the sake of illustration, this quasi-optimal instrumental PDF is compared to the optimal (although impractical) one in Figure \ref{fig:InstrumentalPDFComparison} using the example of Section \ref{sec:ProbabilisticClassificationFunction}.

\subsection{The metamodel-based importance sampling estimator}\label{sec:pf_metaIS}

Choosing the proposed quasi-optimal instrumental PDF (\ie substituting $\widehat{h^*}$ for $h$ in \eqref{eq:pfIS_def}) leads to the following new expression of the failure probability:
\begin{align}
  p_f &= \int \mathbbm{1}_{g \leq 0}(\ve{x}) \frac{f(\ve{x})}{\widehat{h^*}(\ve{x})}\,\widehat{h^*}(\ve{x})\,\di{\ve{x}} \\
      &= p_{f\,\varepsilon} \, \int \frac{\mathbbm{1}_{g \leq 0}(\ve{x})}{\pi(\ve{x})}\,\widehat{h^*}(\ve{x})\,\di{\ve{x}} \\
      &\equiv p_{f\,\varepsilon} \, \alpha_{\rm corr} \\
  \intertext{where we have introduced:}
  \alpha_{\rm corr} &\equiv \Espe{\widehat{h^*}}{\frac{\mathbbm{1}_{g \leq 0}(\ve{X})}{\pi(\ve{X})}} \label{eq:pfcorr}
\end{align}
This means that the failure probability is equal to the product of the augmented failure probability $p_{f\,\varepsilon}$ and a correction factor $\alpha_{\rm corr}$.\par

This correction factor is defined as the expected ratio between the real indicator function $\mathbbm{1}_{g \leq 0}$ and the probabilistic classification function $\pi$. Thus, if the kriging prediction is fully accurate, the correction factor is equal to one and the failure probability is identical to the augmented failure probability (optimality of the proposed estimator). On the other hand, in the more general case where the kriging prediction is not fully accurate (since it is obtained from a DOE $\cx$ of finite size $m$), the correction factor modifies the augmented failure probability accounting for the epistemic uncertainty in the prediction.\par

The two terms in the latter definition of the failure probability may now be estimated using Monte Carlo simulation:
\begin{align}
  \widehat{p}_{f\,\varepsilon} &= \frac{1}{N_{\varepsilon}} \sum\limits_{i=1}^{N_{\varepsilon}} \pi(\ve{x}^{(i)}) \label{eq:pf_epsilon_est}\\
  \widehat{\alpha}_{\rm corr} &= \frac{1}{N_{\rm corr}} \sum\limits_{j=1}^{N_{\rm corr}} \frac{\mathbbm{1}_{g \leq 0}(\ve{h}^{(j)})}{\pi(\ve{h}^{(j)})} \label{eq:alpha_corr_est}
\end{align}%
where the first $N_{\varepsilon}$-sample set is generated from the original PDF $f_{\ve{X}}$, and the second $N_{\rm corr}$-sample set is generated from the quasi-optimal instrumental PDF $\widehat{h^*}$. According to the central limit theorem, these two estimates are unbiased and normally distributed. Their respective variance denoted by $\sigma_{\varepsilon}^2$ and $\sigma_{\rm corr}^2$ may be easily derived as in \eqref{eq:varISest}.\par

Finally, the proposed estimator of the failure probability simply reads as follows:
\begin{equation} \label{eq:pf_metaIS_est}
  \widehat{p}_{f\,{\rm metaIS}} = \widehat{p}_{f\,\varepsilon} \, \widehat{\alpha}_{\rm corr}
\end{equation}%
It is important to note that both terms in \eqref{eq:pf_metaIS_est} are independent, since they rely upon sampling according to two independent PDFs. Based on this latter remark, it is shown in \ref{app:cov_pf_metaIS} that for reasonably small values of the coefficients of variation $\delta_{\widehat{p}_{f\,\varepsilon}}$ and $\delta_{\widehat{\alpha}_{\rm corr}}$ of the two estimators $\widehat{p}_{f\,\varepsilon}$ and $\widehat{\alpha}_{\rm corr}$ (say 1 -- 10\%), the coefficient of variation of the proposed estimator approximately reads:
\begin{equation} \label{eq:cov_metaIS_est}
    \delta_{\widehat{p}_{f\,{\rm metaIS}}} \mathop{\approx}\limits_{\delta_{\varepsilon},\delta_{\rm corr} \ll 1} \sqrt{\delta_{\widehat{p}_{f\,\varepsilon}}^2 + \delta_{\widehat{\alpha}_{\rm corr}}^2}
\end{equation}\par

\subsection{Sampling from the quasi-optimal instrumental density}\label{sec:MCMC}

The instrumental density $\widehat{h^*}$ in \eqref{eq:quasioptISdens} is not straightforward to sample from. First one observes that it is defined up to an unknown but finite normalizing constant:%
\begin{equation}
    \widehat{h^*}(\ve{x}) \propto \pi(\ve{x})\,f(\ve{x})
\end{equation}%
Indeed the exact value of $p_{f\,\varepsilon}$ is not known (it will only be estimated by Monte Carlo simulation). It is clear that the usual \emph{inverse transform} simulation techniques are not applicable for this non-parameteric multidimensional pseudo-PDF. This is the reason why one resorts to \emph{Markov chain Monte Carlo} (MCMC) simulation techniques \citep{Robert2004}.\par

These techniques consists in generating a first-order Markov chain $\cz=\acc{\ve{z}^{(i)}, i=1\enu N}$ whose asymptotic distribution is shown to be the target distribution \citep[see][pp. 206 -- 207]{Robert2004}. The present implementation makes use of the \emph{slice sampling} technique \citep{Neal2003} -- see \ref{app:SliceSampling}.\par

However, \citet{Robert2004} also point out that the Markov chain samples are \emph{not} independent and identically distributed: there is some correlation between the successive states of the chain. Thus, in order to ensure that the samples used in \eqref{eq:pf_epsilon_est} are independent and identically distributed, we use the so-called thining procedure which consists in keeping only one sample every $t$ states of the chain (say $t=10$).\par


\section{Adaptive refinement of the probabilistic classification function}\label{sec:DOE}

The significance of the variance reduction introduced by the proposed metamodel-based importance sampling technique mostly relies on the optimality of the proposed  instrumental PDF $\widehat{h^*}$. Thus, it is proposed here to adaptively refine the probabilistic classification function so that the quasi-optimal instrumental PDF $\widehat{h^*}$ converges towards its optimal counterpart $h^*$.\par

\subsection{Refinement strategy}

\subsubsection{A short state-of-the-art}

There exists a relatively large literature about the adaptive refinement of a kriging prediction for accurate classification (or level-set approximation): see \eg \citet{Oakley2004b,Lee2008,Bichon2008,Vazquez2009,Picheny2010b,Echard2011}. They all rely on the definition of a so-called \emph{in-fill criterion} which is maximum or minimum in the region of interest: the region where the sign of the predicted performance function is the most uncertain. The interested reader may find the expressions for all these criteria together with a discussed comparison on two analytical examples in a recent paper by \citet{Bect2011}.\par

For instance, \citet{Echard2011} propose the so-called \emph{$\cu$-criterion} which is defined as follows:%
\begin{equation}
    \cu(\ve{x}) = \left|\frac{0 - \mu_{\widehat{G}}(\ve{x})}{\sigma_{\widehat{G}}(\ve{x})}\right|
\end{equation}%
This function is minimum in the vicinity of the limit-state surface of the mean prediction $\acc{\ve{x}\in\Rr^n\,:\,\mu_{\widehat{G}}(\ve{x}) = 0}$. This vicinity corresponds to the grayest part of the shade area in Figure \ref{fig:ProbabilisticClassificationFunction_2D}.\par

Then the problem of finding \emph{the best} improvement point is usually reduced to a global optimization problem. Using the $\cu$-criterion from \citet{Echard2011}, the problem reads as follows:%
\begin{equation}
    \ve{x}_{m+1} = \arg \min\limits_{\ve{x} \in \cd_{\ve{X}}} \cu(\ve{x})
\end{equation}%
and is usually solved by means of a numerical global optimization algorithm.\par

Additionally it is argued that for the present reliability estimation problem, one should focus on the most ``probabilistically significant'' region -- such as it is done in FORM, because this is the region where the misclassified points have the most significant impact on the optimality of the proposed estimator. In \citet{Echard2011}, such a trade-off is achieved by picking the best improvement point in a discrete population $\mathfrak{X}$ that is distributed according to $f_{\ve{X}}$. The global optimization is thus reduced to a discrete optimization problem:%
\begin{equation}
    \ve{x}_{m+1} = \arg \min\limits_{\ve{x} \in \mathfrak{X}} \cu(\ve{x})
\end{equation}\par

\subsubsection{The proposed strategy}

As an introduction, it is argued that the available approximation of the optimal instrumental PDF after $m$ observations:%
\begin{equation}
    \widehat{h^*}(\ve{x}) \propto \pi(\ve{x})\,f_{\ve{X}}(\ve{x})
\end{equation}%
is an interesting trade-off criterion because it is maximum \emph{(i)} where the sign of the predicted performance function is supposed to be negative, and \emph{(ii)} where the probability density function $f_{\ve{X}}$ is maximum.\par

Then, it is proposed to use a sampling alternative to the optimization step -- as in \citet{Dubourg2011}. This is achieved by considering the in-fill criteria as PDFs defined up to an unknown but finite normalizing constant. It proceeds as follows:
\begin{enumerate}
    \item Sample a large population (say $N=10^4$ samples) from the weighted in-fill criterion (here $\widehat{h^*}$) using a MCMC simulation technique such as slice sampling \citep{Neal2003}.\label{step1}
    \item Reduce this population to its $K$ clusters' center ($K$ being given, see Section \ref{sec:Appli}) using the $K$-means clustering algorithm \citep{MacQueen1967}.
    \item Evaluate the performance function on the $K$ clusters' center.
    \item Enrich the former experimental design with these $K$ new observations.
    \item Update the kriging prediction and loop back to step \ref{step1} until some target accuracy is achieved -- see Section \ref{sec:stop}.
\end{enumerate}\par

The proposed sampling-based refinement strategy allows one to refine the prediction from a batch of optimal points instead of a single best point. It thus solves the problem of locally optimal points. Indeed, the in-fill criteria proposed in the literature commonly features several \emph{optima} thus meaning that there does \emph{not} exist a single best point. In the proposed strategy all the \emph{maxima} are interpreted as modes of the improper PDF and results in local concentrations of points in the first step. The $K$-means clustering algorithm used in the second step reduces the population generated in the first step to the most significant modes. Note also that this approach allows one to perform several computations of the performance model in a distributed manner (\ie on a high performance computing platform).\par

\subsection{Stopping criterion}\label{sec:stop}

Finally a metric is required to check the optimality of the probabilistic classification function. This metric may then be used as a stopping criterion for the previously detailed refinement strategy thus making it \emph{adaptive}. The metric is based on a cross-validation procedure.\par

Cross-validation techniques are classically used for model selection \citep[see \eg][]{Stone1974}. Basically, the design of experiments $\cx = \acc{\ve{x}_i, i=1\enu m}$ is split into a learning subset $\cx_L$ and a validation subset $\cx_V$ such that $\cx_L \cap \cx_V = \emptyset$ and $\cx_L \cup \cx_V = \cx$. The model is then built using the learning subset $\cx_L$ (hence denoted by $\tilde{g}_{\cx_L}$) and validated by comparing the predicted values $\tilde{g}_{\cx_L}(\ve{x})$ and the real values $g(\ve{x})$ onto the validation subset $\ve{x} \in \cx_V$. The \emph{leave-one-out} technique is a special case where the learning subset is defined as $\cx_L = \cx \setminus \ve{x}_i$. In a regression context, \citet{Allen1971} propose to use a leave-one-out estimate of the mean squared error referred to as the \emph{predicted residual sum of squares}:%
\begin{equation}
    {\rm PRESS} = \frac{1}{m} \sum\limits_{i=1}^m \left(\tilde{g}_{\cx \setminus \ve{x}_i}(\ve{x}_i) - g(\ve{x}_i)\right)^2
\end{equation}\par

The metric proposed in this paper is a leave-one-out estimate of the correction factor in \eqref{eq:pfcorr}:%
\begin{align} \label{eq:corr_LOO}
  \widehat{\alpha}_{\rm corr\,LOO} = \frac{1}{m} \sum\limits_{i = 1}^m \frac{\mathbbm{1}_{g \leq 0}(\ve{x}_i)}{\cp\left[\widehat{G}_{\cx \setminus \ve{x}_i}(\ve{x}_i) \leq 0\right]}
\end{align}%
where $\widehat{G}_{\cx \setminus \ve{x}_i}$ is the $i$-th leave-one-out kriging prediction of the performance function built from the DOE $\cx$ without the $i$-th sample $\ve{x}_i$. This quantity should be estimated using a minimal number of samples (say $m = 30$) so that it is sufficiently accurate.\par

The reason for introducing this leave-one-out estimate of $\widehat{\alpha}_{\rm corr}$ is the following. In the early steps of the refinement procedure, the kriging predictor is not accurate enough to evaluate the failure probability by means of \eqref{eq:pf_metaIS_est} efficiently. It would thus be ineffective to waste costly evaluations of the true limit-state function to compute the true correction factor in \eqref{eq:alpha_corr_est}. On the contrary, the leave-one-out estimate $\widehat{\alpha}_{\rm corr\,LOO}$ makes only use of the \emph{available observations} in the experimental design, so that it is fast to evaluate. As discussed earlier in Section \ref{sec:pf_metaIS}, a sufficient accuracy is reached when the estimate in \eqref{eq:corr_LOO} gets close to one because it means that the probabilistic classification function converges towards its deterministic counterpart. And consequently the instrumental PDF $\widehat{h^*}$ converges towards it optimal counterpart $h^*$. It is thus proposed to use $\widehat{\alpha}_{\rm corr\,LOO}$ as an indication to stop the refinement strategy. Precisely, the iterative enrichment of $\cx$ is stopped if its size is greater than 30 and $\widehat{\alpha}_{\rm corr\,LOO}$ is in the order of magnitude of 1. In the case of high dimensional and/or highly nonlinear problems, the size of the DOE is limited to $m_{\rm max} = 1\,000$. One may also measure the improvement brought by the latest refinement iteration in terms of variation of $\widehat{\alpha}_{\rm corr\,LOO}$.\par

\paragraph{Remark regarding numerical implementation} It may happen that a leave-one-out estimate of the kriging variance $\sigma_{\widehat{G}_{\cx \setminus \ve{x}_i}}^2(\ve{x}_i)$ gets close or even equal to zero. This is problematic because the ratio in \eqref{eq:corr_LOO} might not be defined in such cases. Indeed, if the mean prediction $\mu_{\widehat{G}_{\cx \setminus \ve{x}_i}}(\ve{x}_i)$ is positive, the probabilistic classification function in the denominator equals zero -- and it may cause an exception. It is argued that this variance should never be exactly zero (the kriging variance equals zero only at the samples in the DOE, here $\cx \setminus \ve{x}_i$), so that it is proposed to bound the probabilistic classification function above a reasonably low value (say the machine precision, $\varepsilon_{\rm M} \approx 10^{-16}$).


\section{Algorithmic implementation} \label{sec:Implementation}

\begin{figure}
    \includegraphics[width=\textwidth]{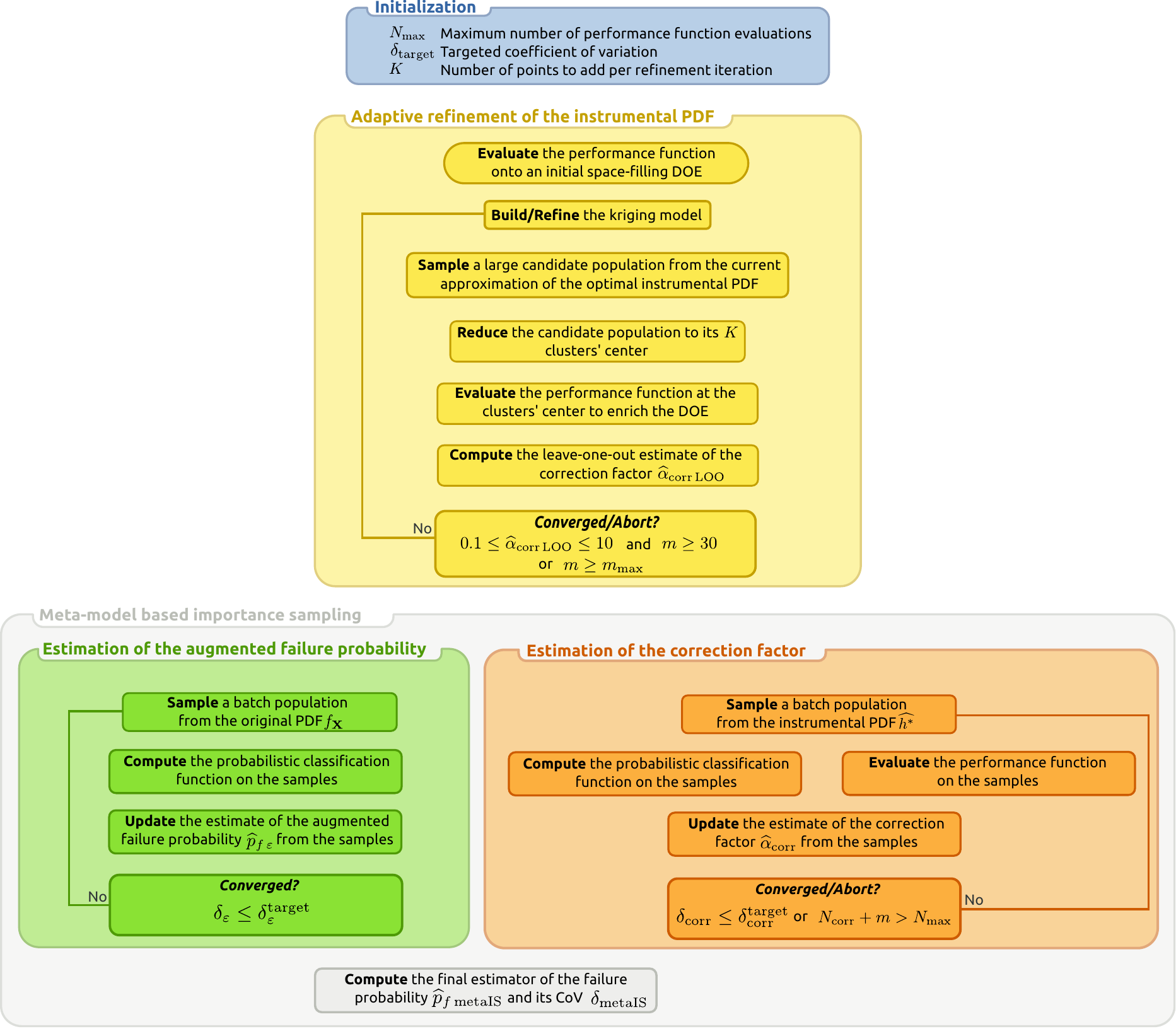}
    \label{fig:MetamodelBasedImportanceSampling}
    \caption{Flowchart of the proposed metamodel-based importance sampling algorithm}
\end{figure}

The purpose of this section is to summarize the proposed metamodel-based importance sampling algorithm. The flowchart of the algorithm is given in Figure \ref{fig:MetamodelBasedImportanceSampling}. It is essentially divided in three independent steps, two of which could potentially be run in parallel.\par

First, the algorithm only requires the choice of a maximum number of performance function evaluation $N_{\rm max}$, a targeted coefficient of variation $\delta_{\rm target}$ for the final estimate $\widehat{p}_{f\,{\rm metaIS}}$ and the number $K$ of points that should be added at each refinement iteration.\par

Then, the first step of the algorithm is to build a reasonably accurate approximation of the probabilistic classification function that will bring a significant variance reduction for the final importance sampling estimate. Note that this step is optional in the case the analysis is resumed from an existing DOE.\par

The two other steps are independent and might be run in parallel. They consist in using Monte Carlo simulation for the estimation of the two terms $\widehat{\alpha}_{\rm corr}$ and $\widehat{p}_{f\,\varepsilon}$ defining the proposed estimator $\widehat{p}_{f\,{\rm metaIS}}$ -- see Section \ref{sec:metaIS}. In the current implementation, we targeted the same coefficient of variation $\delta_{\rm \varepsilon}=\delta_{\rm corr}=\delta_{\rm target}/\sqrt{2}$ for the two estimators. It could also be possible to implement these two steps in two independent threads that would communicate between each other and decide whether to stop the simulation or not based on the estimate of the final coefficient of variation. It should also be pointed out that it is easier to reduce the estimation variance on the augmented failure probability than the one one the correction factor because the former only depends on the kriging surrogate.\par


\section{Application examples} \label{sec:Appli}

In this section, the proposed strategy is applied to two structural reliability problems. The first one involves a simple analytical limit-state function and may thus be used to validate the implementation of the proposed strategy whereas the second example involves a more sophisticated nonlinear finite element model coupled with a high dimensional stochastic model so as to prove the applicability of the strategy to industrial problems.

\subsection{Analytical limit-state function -- Influence of the dimension}

This basic structural reliability example is taken from \citet{Rackwitz2001}. It involves $n$ independent lognormal random variates with mean value $\mu=1$ and standard deviation $\sigma=0.2$. The performance function reads as follows:%
\begin{equation}
 g(\ve{x}) = \left(n + a\,\sigma\,\sqrt{n}\right) - \sum\limits_{i=1}^n x_i
\end{equation}%
where $a$ is set equal to 3 for the present application. The dimension of the problem is successively set equal to $n=\acc{2,\,50,\,100}$ to assess the influence of the dimension on the proposed estimator. The results are given in Table~\ref{tab:Rackwitz}.\par

\begin{table}[H]
  \begin{center}
    \begin{footnotesize}
      \renewcommand{\arraystretch}{1.2}
      \begin{tabular}{cccc}
        \hlineT
                        $n$             &           2         &          50         &         100         \\
        \hlineB
        \multicolumn{4}{l}{\cellcolor[gray]{.9}\textbf{Crude Monte Carlo (ref.)}} \\
        $\widehat{p}_{f\,{\rm MC}}$     & $4.78\times10^{-3}$ & $1.91\times10^{-3}$  & $1.73\times10^{-3}$ \\
        $\delta_{\rm MC}$               &     $\leq 2\%$      &     $\leq 2\%$       &     $\leq 2\%$      \\
        $N$                             &       522,000       &      1,100,000       &    1,450,000      \\
        \hlineB
        \multicolumn{4}{l}{\cellcolor[gray]{.9}\textbf{Metamodel-based importance sampling}} \\
        $N_{\rm DOE}$                   &      $6\times2$     &     $6\times50$     &     $6\times100$    \\
        $\widehat{p}_{f\,\varepsilon}$  & $5.03\times10^{-3}$ & $1.95\times10^{-3}$ & $1.83\times10^{-3}$ \\
        $\delta_{\varepsilon}$          &    $\leq 1.41\%$    &    $\leq 1.41\%$    &    $\leq 1.41\%$    \\
        \hline
        $N_{\rm corr}$                  &         100         &        1,500        &       2,100         \\
        $\widehat{\alpha}_{\rm corr}$   &          1          &        0.99         &        0.93         \\
        $\delta_{\rm corr}$             &         0\%         &    $\leq 1.41\%$    &    $\leq 1.41\%$    \\
        \hline
        $N_{\rm DOE} + N_{\rm corr}$    &         112         &        1,800        &       2,700         \\
        $\widehat{p}_{f\,{\rm metaIS}}$ & $5.03\times10^{-3}$ & $1.93\times10^{-3}$ & $1.70\times10^{-3}$ \\
        $\delta_{\rm metaIS}$           &    $\leq 1.41\%$    &     $\leq 2\%$      &     $\leq 2\%$      \\
        \hlineB
      \end{tabular}
    \end{footnotesize}
  \end{center}
  \caption{Comparative results for the performance function $g(\ve{x}) = \left(n + a\,\sigma\,\sqrt{n}\right) - \sum\limits_{i=1}^n x_i$ from \citet{Rackwitz2001}.}
  \label{tab:Rackwitz}
\end{table}

The first block provides the results from a crude Monte Carlo simulation (reference) whereas the second block provides the results from the proposed importance sampling scheme. For both methods a 2\% coefficient of variation on the failure probability estimates is targeted. First, it can be observed that the proposed estimator is in good agreement with the reference Monte Carlo estimates. The estimates of both the augmented failure probability $\widehat{p}_{f\,\varepsilon}$ and the correction factor $\widehat{\alpha}_{\rm corr}$ are also provided. Note that the correction factor increases with the dimension. In low dimension ($n=2$) the kriging surrogate almost exactly equals the performance function (at least in the vicinity of its limit-state) hence $\widehat{\alpha}_{\rm corr}=1$ and $\widehat{p}_{f\,{\rm metaIS}} = \widehat{p}_{f\,\varepsilon}$ (no misclassification). In larger dimensions the surrogate loses accuracy, and the correction factor gets more and more influent.\par

The size of the DOE from which the kriging prediction is built is given as the product between the number of refinement iterations and the number $K$ of clusters' center added per iteration. $K$ is chosen equal to the number of input random variables ($n=\acc{2,\,50,\,100}$) so that the cost for the definition of the instrumental PDF is equivalent to the cost induced by a gradient-based design point search algorithm (when the gradient is approximated using a \emph{finite forward differentiation} technique). In all cases 6 iterations are required (see Table \ref{tab:Rackwitz}). Note that the total number of calls to the real performance function $g$ for the proposed importance sampling scheme ($N_{\rm DOE} + N_{\rm corr}$) is much less than the number of calls induced by Monte Carlo simulation ($N$) for the same targeted coefficient of variation.

\subsection{Nonlinear stability analysis of an imperfect shell roof}

\begin{figure}
    \centering
    \includegraphics[width=.6\linewidth, clip=true, trim=0 -30 0 -30]{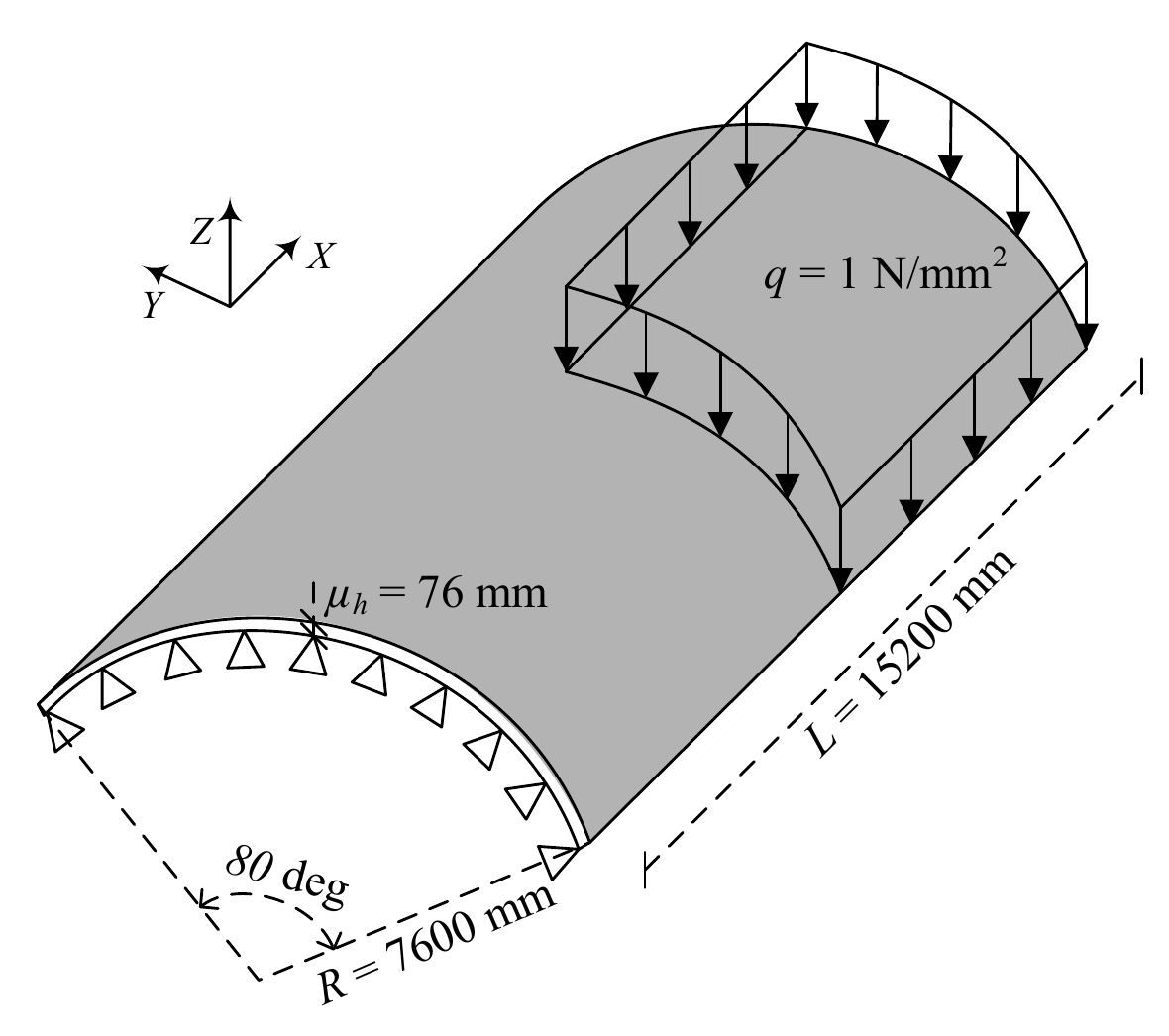}
    \caption{Illustration of the Scordelis-Lo shell roof example.}
    \label{fig:ScordelisLo}
\end{figure}

The mechanical model for this example is inspired by \citet{Scordelis1961}. It concerns the buckling analysis of a cylindrical shell roof whose dimensions are given in Figure \ref{fig:ScordelisLo}. The longitudinal edges of the roof are free while its circumferential edges are supported by rigid diaphragms (radial displacement fixed to zero). Its constitutive material is assumed to have a nonlinear elastic Ramberg-Osgood material behavior. It is subjected to a constant surface load $q$ and the structure is considered to fail if its critical buckling load $q_{\rm cr}$ is less than a prescribed service load of magnitude $q_0=0.18$~MPa, so that the associated performance function may be defined as follows:
\begin{equation}
  g(\ve{\xi}) = q_{\rm cr}(\ve{\xi}) - q_0
\end{equation}%
where $\ve{\xi}$ denotes the outcome of the random vector $\ve{\Xi}$ introduced in the sequel.\par

The critical buckling load $q_{\rm cr}$ is determined by means of the \emph{asymptotic numerical method} \hbox{\citep{Cochelin1994a}} coupled with a $30\times30$ 8-node Büchter-Ramm shell finite element mesh \citep{Buchter1994} using the EVE finite element code. This academic software was initially developed by \hbox{\citet{Cochelin1994a}} and further developed by \citet{NoirfaliseWCCM2008}.\par

The stochastic model, inspired from \citet{DubourgIcossar2009}, involves four independent random fields defined over the roof surface. They describe respectively: the initial shape imperfection $\zeta$, the material Young's modulus $E$, the material yield stress $\sigma_y$, and the shell thickness $h$. The random shape imperfection is modeled as a linear combination of the three most critical Euler buckling modes' shape $\acc{{\cal U}_k,\,k=1\enu 3}$, so that it reads as follows:
\begin{equation}
  \zeta(\ve{x},\,\theta) = \sum\limits_{k=1}^3 \Xi_{\zeta\,k}\,{\cal U}_k(x,\,\theta)
\end{equation}
where $\acc{\Xi_{\zeta\,k}, k=1\enu 3}$ are three independent Gaussian random variates with mean $\mu_{\zeta} = 0$ and standard deviation $\sigma_{\zeta} \approx 9.5$~mm. This empirical model was determined with the following objectives:
\begin{itemize}
    \item the shape imperfection is Gaussian with a zero mean, so that the random variables should also be Gaussian with a zero mean;
    \item the maximum amplitude of the shape imperfection around the mean surface ($\mu_{\zeta} = 0$) $\pm 2\,\sigma_{\zeta}$ is set equal to the half of the mean thickness $\mu_h = 76$~mm, leading to $\sigma_{\zeta} \approx 9.5$~mm.
\end{itemize}\par

The other three random fields are assumed independent and lognormal with constant means $\mu_h = 76$~mm, $\mu_E = 200,000$~MPa and $\mu_{\sigma_y} = 390$~MPa and coefficients of variation $\delta_h = 5\%$, $\delta_E = 3\%$ and $\delta_{\sigma_y} = 7\%$. They are represented by three Karhunen-Loève expansions of three independent standard Gaussian random fields, whose sample paths are \emph{translated} into lognormal sample paths -- see \citet{DubourgIcossar2009} for the mapping. These three Gaussian random fields are assumed to have the same isotropic squared exponential autocorrelation function with a correlation length $\ell = 3,500$~mm.\par

Due to the choice of this correlation function, the Fredholm integral equation involved in the Karhunen-Loève discretization scheme has no analytical solution. A so-called \emph{wavelet-Galerkin} numerical procedure was thus used as detailed in \citet{Phoon2002}. Each random field is simulated by means of 30 independent standard Gaussian random variates leading to a relative mean squared discretization error of 3.70\%. Finally the complete stochastic model involves 93 independent random variables.\par

\begin{table}
  \begin{center}
    \begin{footnotesize}
      \renewcommand{\arraystretch}{1.2}
      \begin{tabular}{cccc}
        \hlineT \rowcolor[gray]{.9}
        \textbf{Method}          &  \textbf{Subset (ref.)}  &  \textbf{Multi-FORM}  &  \textbf{Meta-IS}   \\
        \hlineT
        \textbf{DOE size}        &            -             &           -           &    $6\times93$      \\
        \textbf{MPFP search}     &            -             &    $\approx10,000$    &         -           \\
        \textbf{Simulations}     &         $20,000$         &           -           &      $9,464$        \\
        \textbf{$p_f$ estimates} &   $1.27\times10^{-4}$    &  $1.22\times10^{-4}$  & $1.32\times10^{-4}$ \\
        \textbf{C.o.V.}          &         12.36\%          &           -           &       13.75\%       \\
        \hlineB
        \end{tabular}
      \end{footnotesize}
    \end{center}
    \caption{Comparative results for the shell roof example adapted from \citet{DubourgIcossar2009}.}
  \label{tab:ScordelisLo}
\end{table}

The reliability results are gathered in Table \ref{tab:ScordelisLo}. The proposed importance sampling scheme leads to a failure probability in full agreement with the value obtained by subset simulation \citep[as implemented in the FERUM toolbox v4.0 by][]{Bourinet2009} which validates the proposed strategy. For this example the augmented failure probability $\widehat{p}_{f\,\varepsilon}$ is equal to $2.06\times10^{-4}$ (with a coefficient of variation of 5.70\%), and the correction factor $\widehat{\alpha}_{\rm corr}$ is equal to $0.641$ (with a coefficient of variation of 12.49\%).\par

\begin{figure}
    \centering
    \begin{tikzpicture}[scale=1.8]
        \node (N1) at (0,0) {\includegraphics[width=\textwidth, clip=true, trim=50 50 50 50]{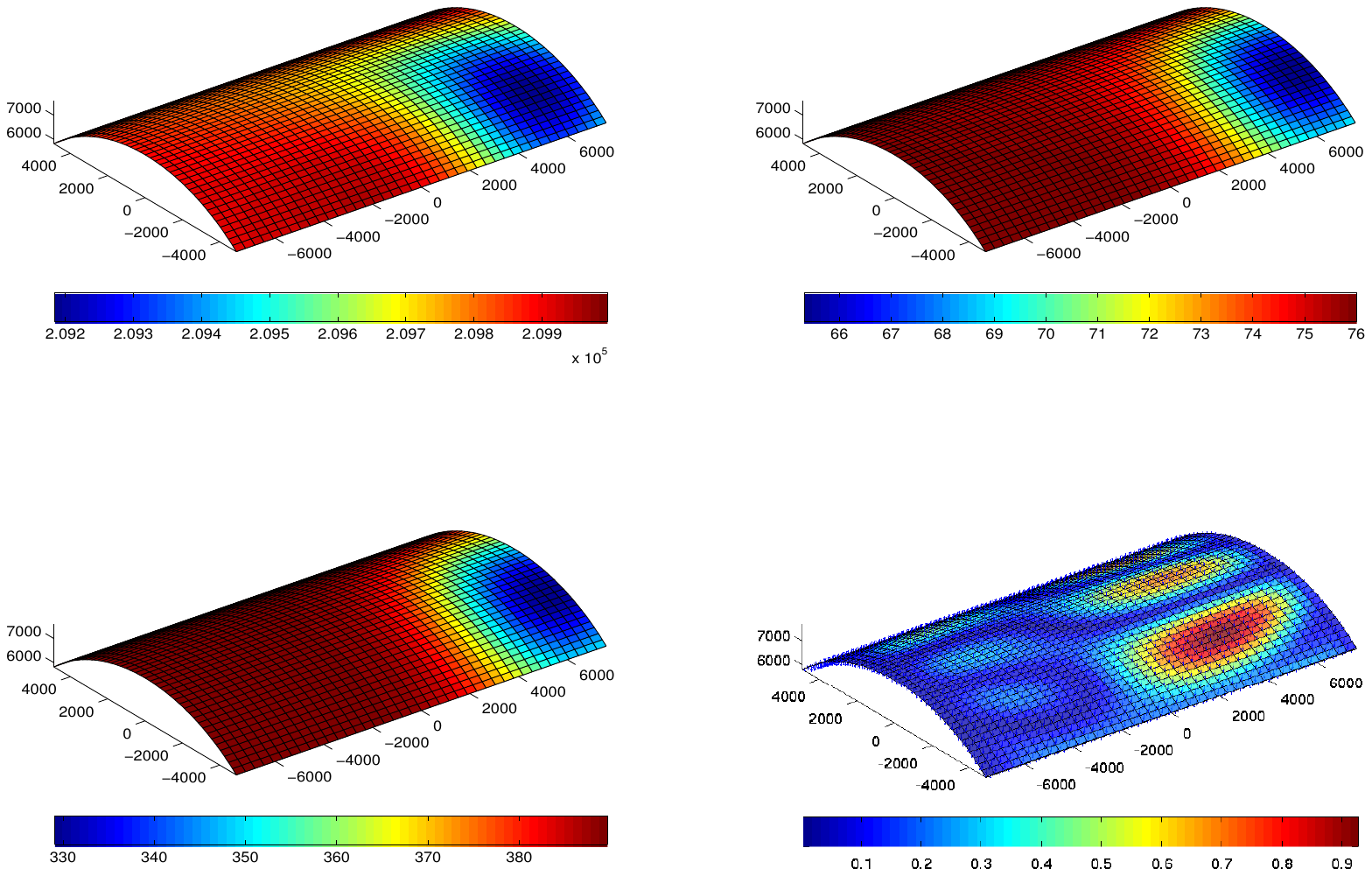}};
        \node [rectangle] (N2) at (-3.2,2.7) {$E^{*\,(4)}$};
        \node [rectangle] (N2) at (-3.2,-.7) {$\sigma_y^{*\,(4)}$};
        \node [rectangle] (N2) at (1.4,2.7) {$h^{*\,(4)}$};
        \node [rectangle] (N2) at (1.4,-.7) {$\zeta^{*\,(4)}$};
    \end{tikzpicture}
    \caption{One of the 4 most probable failure configurations.}
    \label{fig:ScordelisLo_MPFP}
\end{figure}

A multiple FORM analysis revealed the existence of 4 most probable failure configurations identified by means of the restarted \emph{i}-HLRF algorithm from \citet{DerKiureghian1998}. These 4 failure modes corresponds to the 4 extreme cases for which the ``demand'' random field $\zeta$ is maximal in the corner of the roof whereas the ``capacity'' random fields $E$, $\sigma_y$ and $h$ are minimal -- one of these configurations is illustrated in Figure~\ref{fig:ScordelisLo_MPFP}. The combination of these 4 failure modes in a serial system then allowed to give a \emph{Multi-FORM} approximation of the failure probability (third column in Table~\ref{tab:ScordelisLo}). In this case, the Ditlevsen's bounds \citep{Dit2} coincide (up to the accuracy provided in Table~\ref{tab:ScordelisLo}). It thus proves the ability of the proposed strategy to deal with reliability problems featuring multiple design points in a reasonably high dimension.\par

\section{Conclusion}

Starting from the double premise that the usual surrogate-based reliability analyses do not permit to quantify the error made by using the metamodel instead of the original limit-state function, and that the existing variance reduction techniques remain time-consuming when the performance function involves the output of an expensive-to-evaluate black box function, an hybrid strategy has been proposed.\par

First, a probabilistic classification function based on the post-processing of a kriging prediction was introduced. This function allows a smoother classification than its deterministic counterpart (\ie the indicator function of the failure domain) accounting for the epistemic uncertainty in the kriging prediction. The probabilistic classification is then used to formulate a quasi-optimal importance sampling density. Using elementary algebra the failure probability is recast as a product of two terms, namely the \emph{augmented failure probability} $p_{f\,\varepsilon}$ which is evaluated by means of the meta-model only, and a correction factor $\alpha_{\rm corr}$ that is computed from evaluations of the original limit-state function. In order to decide whether the kriging surrogate is accurate enough, a leave-one-out estimate of $\alpha_{\rm corr}$ is used and the iterative refinement is stopped when it is in the order of magnitude of 1. Once the kriging surrogate has been built, the two terms of the product defining the failure probability may be evaluated in parallel.\par

The method turned out to be efficient on several application examples, two of which are presented in this paper. It can handle problems featuring a reasonably high number of random variables and multiple design points. Further work is in progress to include the proposed algorithm within a reliability-based design optimization framework.\par


\section*{Acknowledgements}\addcontentsline{toc}{section}{Acknowledgements}

The first author is funded by a CIFRE grant from Phimeca Engineering S.A. subsidized by the ANRT (convention number 706/2008). The financial support from the French National Research Agency (ANR) through the KidPocket project is also gratefully acknowledged.


\appendix
\section{Calculation of the coefficient of variation}\label{app:cov_pf_metaIS}

The estimator in \eqref{eq:pf_metaIS_est} is defined as the product of two estimators. Let us denote these two unbiased independent estimators as $\widehat{p_1}$, $\widehat{p_2}$ with variances $\sigma_1^2$, $\sigma_2^2$ for the sake of clarity. The calculation of the variance of the final estimator $\widehat{p} = \widehat{p_1}\,\widehat{p_2}$ proceeds as follows.\par

First, according to its definition, the variance reads:
\begin{equation}
  \sigma_{\widehat{p}}^2 \equiv \Var{\widehat{p}} \equiv \Esp{\widehat{p_1}^2\,\widehat{p_2}^2} - \Esp{\widehat{p_1}\,\widehat{p_2}}^2
\end{equation}%
Since the two estimators $\widehat{p_1}$ and $\widehat{p_2}$ are independent, the variance also reads:
\begin{equation}
  \sigma_{\widehat{p}}^2 = \Esp{\widehat{p_1}^2}\,\Esp{\widehat{p_2}^2} - \Esp{\widehat{p_1}}^2\,\Esp{\widehat{p_2}}^2
\end{equation}%
which may be further elaborated according to the König-Huyghens theorem:
\begin{equation}
    \sigma_{\widehat{p}}^2 = \left(\Esp{\widehat{p_1}}^2 + \sigma_1^2\right)\,\left(\Esp{\widehat{p_2}}^2 + \sigma_2^2\right) - \Esp{\widehat{p_1}}^2\,\Esp{\widehat{p_2}}^2
\end{equation}%
Due to the unbiasedness of the estimators, one finally gets:
\begin{eqnarray}
  \sigma_{\widehat{p}}^2 & = & \left(p_1^2 + \sigma_1^2\right)\,\left(p_2^2 + \sigma_2^2\right) - p_1^2\,p_2^2 \\
                         & = & \sigma_1^2\,\sigma_2^2 + p_1^2\,\sigma_2^2 + p_2^2\,\sigma_1^2
\end{eqnarray}

Denoting by $\delta_i=\sigma_i/p_i$ the coefficients of variation of $\widehat{p_i},\,i=1,\,2$, the coefficient of variation of $\widehat{p} = \widehat{p_1}\,\widehat{p_2}$ eventually reads:
\begin{equation}
  \delta \equiv \frac{\sigma_{\widehat{p}}}{p_1\,p_2} = \sqrt{\delta_1^2 + \delta_2^2 + \delta_1^2\,\delta_2^2}
\end{equation}

In practice usual target coefficients of variation $\delta_{\rm target}$ range from 1\% to 10\% so that:
\begin{equation}
    \delta \mathop{\approx}\limits_{\delta_1,\delta_2 \ll 1} \sqrt{\delta_1^2 + \delta_2^2}
\end{equation}

\section{The slice sampling technique}\label{app:SliceSampling}

\begin{figure}
    \centering
    \includegraphics[width=.8\columnwidth, clip=true, trim=20 40 40 25]{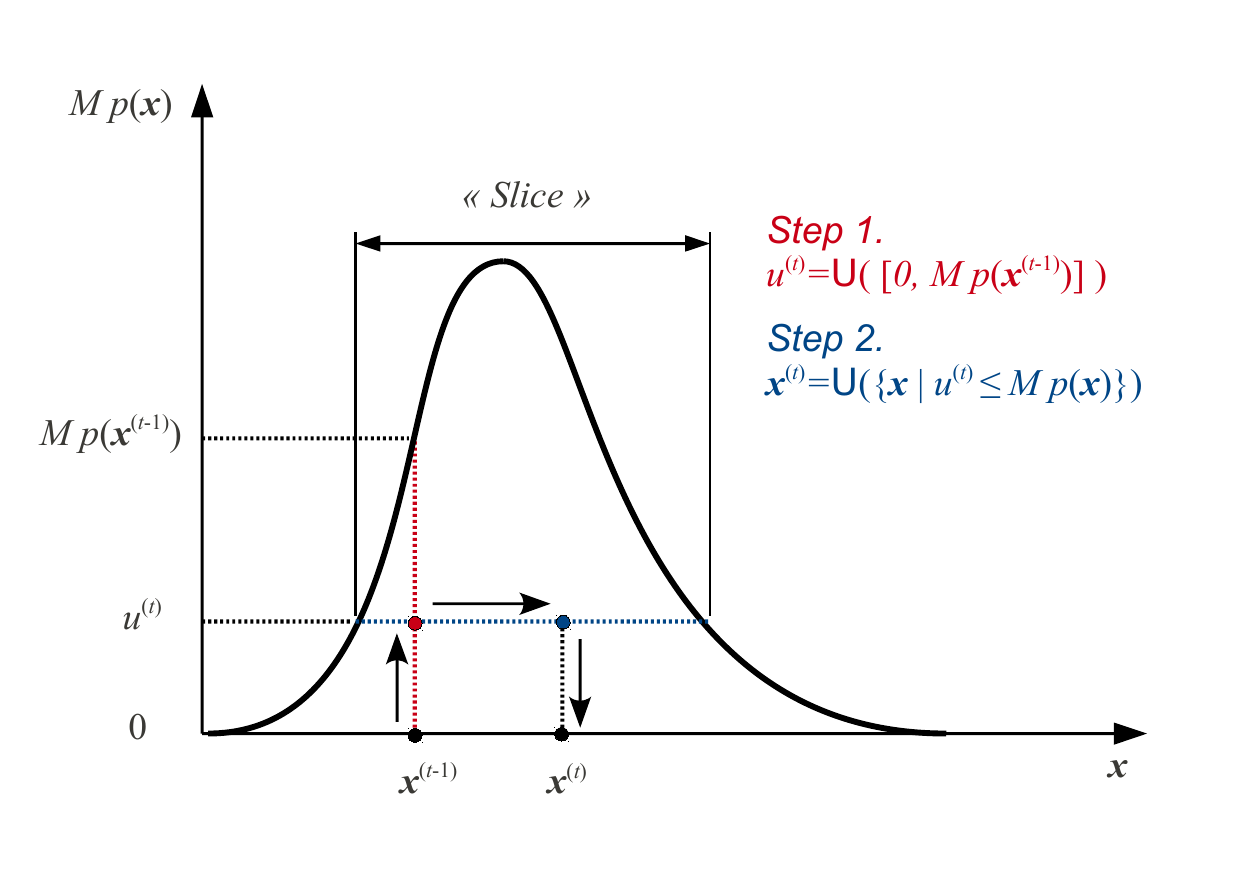}
    \caption{Illustration of the \emph{slice sampling} technique: uniform generation under the target PDF $p$.}
    \label{fig:SliceSampling}
\end{figure}

This simulation technique consists in introducing an auxiliary scalar random variate $U$ such that the joint PDF of the augmented vector $(U, \ve{X})$ reads:
\begin{equation}
    J(u, \ve{x}) = \left\{\begin{array}{rl}
                       1 & \text{if}\quad0 \leq u \leq p(\ve{x}) \\
                       0 & \text{otherwise}
                   \end{array}\right.
\end{equation}%
Note that this definition is nothing but a translation of the fundamental theorem of simulation \citep[see][Theorem 2.15]{Robert2004} whose key idea is to sample the augmented vector ``under'' the density curve (or surface) $p(\ve{x})$ -- see also Figure \ref{fig:SliceSampling}.\par

It is easy to show that the \emph{marginalization} of $J(u, \ve{x})$ \emph{w.r.t.} $u$ results in the target density $p$:
\begin{equation}
    \int_{\Rr} J(u, \ve{x})\,\di{u} = \int_0^{p(\ve{x})} 1\,\di{u} = p(\ve{x})
\end{equation}
Hence, sampling $\ve{X}$ from $p$ is equivalent to sampling $(U, \ve{X})$ from $J$ and then ignoring $U$.\par

The slice sampling procedure relies on the conditional distributions derived from $J$ which read:%
\begin{eqnarray}
    U \mid \ve{X}=\ve{x} & \sim & \cu\left([0, p(\ve{x})]\right) \\
    \ve{X} \mid U=u & \sim & \cu\left(\cs = \acc{\ve{x} \in \Rr^n: p(\ve{x}) \geq u}\right)
\end{eqnarray}%
The slice $\cs$ may not be easy to identify in high dimensions, but \citet{Neal2003} proposed interesting algorithmic improvements which make the approach scalable to a wide variety of pseudo-PDFs.\par

The two basic steps of the slice sampling algorithm are illustrated in Figure \ref{fig:SliceSampling}. As other Markov chain sampling techniques, it does not require the explicit knowledge of the normalizing constant $M$ of the target PDF $p$ so that the value of $M\,p(\ve{x})$ is sufficient. At iteration $t$, starting from a seed $\ve{x}^{(t-1)}$ that is distributed according to the target PDF (\ie such that $M\,p(\ve{x}^{(t-1)}) \neq 0$), one can first \emph{(i)} generate a random number $u^{(t)}$ which is uniformly distributed between 0 and $M\,p(\ve{x}^{(t)})$, and then \emph{(ii)} generate a random vector $\ve{x}^{(t)}$ that has uniform distribution on the $t$-th slice ${\cs^{(t)} = \acc{\ve{x} \in \Rr^n: M\,p(\ve{x}) \geq u^{(t)}}}$.\par

\section*{References}\addcontentsline{toc}{section}{References}
\bibliographystyle{elsarticle-harv}


\end{document}